\documentclass[11pt,a4paper]{article}
\usepackage{jheppub}

\usepackage[utf8]{inputenc}

\usepackage{graphicx}
\usepackage{amsmath}
\usepackage[toc]{appendix}

\usepackage{xfrac} 

\numberwithin{equation}{section}
\newcommand{\SEC}[1]{Sec.~\ref{sec:#1}}

\newcommand{\VEC}[1]{\boldsymbol{#1}}
\newcommand{\SUM}[2]{\sum\limits_{#1}^{#2}}
\newcommand{\INT}[2]{\int_{#1}^{#2}}
\newcommand{\FEYNINT}[1]{\INT{0}{1}\DIFFL #1}
\newcommand{\DIFFL}{\mathop{}\!\mathrm{d}}
\newcommand{\DIFFD}{\mathop{}\!\mathrm{d}^d}
\newcommand{\DIFFFnopi}[2]{\mathop{}\!\mathrm{d}^{#1}#2} 
\newcommand{\PROD}[2]{\prod\limits_{#1}^{#2}}
\newcommand{\DIFFSPinopi}[2]{\mathop{}\!\mathrm{d}\Omega_{#2}^{(#1)}} 

\newcommand{\DD}[1]{\mathop{}\!\delta\!\left(#1\right)}
\newcommand{\DDP}[1]{\mathop{}\!\delta^+\!\left(#1\right)}
\newcommand{\GF}[1]{\mathop{}\!\Gamma\left(#1\right)}

\newcommand{\GENHYPGF}[5]{\mathop{}\!_{#1}F_{#2}\left[\{#3\},\{#4\};#5\right]}
\newcommand{\HYPGF}[4]{\GENHYPGF{2}{1}{#1,#2}{#3}{#4}}
\newcommand{\GPL}[2]{\text{G}(\{#1\};#2)}
\newcommand{\Li}[2]{\mathop{}\!{\rm Li}_{#1}(#2)}
\newcommand{\Ln}[1]{\mathop{}\!{\rm ln}(#1)}
\newcommand{\Lnp}[2]{\mathop{}\!{\rm ln}^{#1}(#2)}
\newcommand{\xLi}[2]{\mathop{}\!{\rm Li}_{#1}\left(#2\right)}
\newcommand{\xLn}[1]{\mathop{}\!{\rm ln}\left(#1\right)}
\newcommand{\xLnp}[2]{\mathop{}\!{\rm ln}^{#1}\left(#2\right)}

\newcommand{\DS}{S\hspace{-5pt}S}
\newcommand{\ep}{\epsilon}
\newcommand{\ORD}[1]{\mathop{}\!\mathcal{O}\!\left(#1\right) }

\newcommand{\Emax}{E_{\rm{max}}}
\newcommand{\scr}[2]{(#1\!\cdot\! #2)}

\newcommand{\Dcut}{D_{\mathrm{cut}}}
\newcommand{\Imast}[1]{\bigg\langle #1 \bigg\rangle}
\newcommand{\Svol}[1]{\Omega^{(#1)}}
\newcommand{\Tcan}{\hat{T}_{\rm can}}

\newcommand{\G}[1]{\mathcal{G}_{#1}}
\newcommand{\GG}[1]{\mathcal{GG}_{#1}}
\newcommand{\QQ}[1]{\mathcal{Q\bar{Q}}_{#1}}
\newcommand{\ffx}[1]{f^{gg,q\bar{q}}_{#1}}
\newcommand{\ggx}[1]{f^{gg}_{#1}}
\newcommand{\qqx}[1]{f^{q\bar{q}}_{#1}}
\newcommand{\eXX}[1]{\mathcal{E}^{\mathcal{XX}}_{#1}}
\newcommand{\eGG}[1]{\mathcal{E}^{\mathcal{GG}}_{#1}}
\newcommand{\eGGso}[1]{\mathcal{E}^{\mathcal{GG},\mathrm{s.o.}}_{#1}}
\newcommand{\eQQ}[1]{\mathcal{E}^{\mathcal{Q\bar{Q}}}_{#1}}

\author[a,b]{Wojciech Bizo\'{n},}
\author[a]{Maximilian~Delto}

\affiliation[a]{Institut f{\"u}r Theoretische Teilchenphysik (TTP), KIT, 76128 Karlsruhe, Germany}
\affiliation[b]{Institut f{\"u}r Kernphysik (IKP), KIT, 76344 Eggenstein-Leopoldshafen, Germany}

\emailAdd{maximilian.delto@kit.edu}
\emailAdd{wojciech.bizon@kit.edu}

\title{
  Analytic double-soft integrated subtraction terms for two massive emitters in a back-to-back kinematics
}

\preprint{
  \\\\
  TTP20-014 \\
  P3H-20-011
}

\abstract{
  We consider the double-soft limit of QCD amplitudes with two massive
  quarks in a back-to-back kinematics accompanied by two soft
  partons.
  We integrate analytically the respective double-soft eikonal
  functions over the phase space of the two soft partons.
  Within the context of the nested soft-collinear subtraction scheme,
  our results may serve as one of the integrated subtraction terms
  needed for the analytic and fully-differential description of
  next-to-next-to-leading order (NNLO) QCD corrections to
  colour-singlet decay into massive partons or to heavy-quark pair
  production.
}

\usepackage{hyperref}
\allowdisplaybreaks
\begin{document} 

\maketitle
\flushbottom

\section{Introduction}
\label{seq_intro}
The focus of the physics program at the LHC has recently shifted from
direct searches for new particles to precision studies of various
Standard Model (SM) processes.
Such studies are indispensable since, despite the discovery of the
Higgs boson in 2012~\cite{Aad:2012tfa,Chatrchyan:2012ufa} that
formally completed the SM of particle physics, there is a number of
intriguing questions that cannot be answered within this theory.
Given the lack of direct evidence for new particles in collider
experiments, it becomes important to stress-test predictions of the SM
with an unprecedented precision, which becomes possible thanks to the
upcoming high-luminosity phase at the LHC.
As a consequence, high-precision theoretical predictions for many
observables that can be studied in various SM processes at the LHC
become necessary.

The perturbative description of hard scattering processes at the LHC
has to overcome two main obstacles.
One is the computation of multi-scale virtual amplitudes, where loop
integrals over momenta of virtual particles need to be calculated. The
other obstacle is the appearance of infrared singularities during
phase-space integration of real corrections when one or more emissions
become soft or collinear to other partons.

At next-to-leading order (NLO) in perturbative QCD, the treatment of
infrared singularities was tackled long time ago with two generic
methods, \textit{slicing}~\cite{Fabricius:1981sx} and
\textit{subtraction}~\cite{Ellis:1980wv}.
Since nowadays, both virtual and real corrections at NLO can be
calculated in a fully automated way, the applicability of these
methods is limited by computing power only.
The situation changes at the next-to-next-to-leading order (NNLO),
where it is still being debated how to extend the well-established
NLO subtraction schemes~\cite{Frixione:1995ms,Frixione:1997np,Catani:1996vz,Catani:2002hc}
to the next order.

Currently, theoretical predictions with NNLO QCD accuracy exist for
many LHC processes. They were obtained using slicing methods that
include
$q_T$-~\cite{Catani:2007vq,Bonciani:2015sha,Grazzini:2017mhc,Catani:2019hip}
and
$N$-jettiness~\cite{Boughezal:2015dva,Boughezal:2015aha,Gaunt:2015pea,Boughezal:2016wmq}
slicing; as well as subtraction schemes such as antenna subtraction
\cite{GehrmannDeRidder:2005cm,GehrmannDeRidder:2005aw,GehrmannDeRidder:2005hi,Daleo:2006xa,Daleo:2009yj,Gehrmann:2011wi,Boughezal:2010mc,GehrmannDeRidder:2012ja,Currie:2013vh,Currie:2017eqf,Currie:2018xkj},
geometric subtraction~\cite{Herzog:2018ily}, the STRIPPER framework
\cite{Czakon:2010td,Czakon:2011ve,
  Czakon:2014oma,Czakon:2019tmo,Chawdhry:2019bji}, local analytic
sector subtraction~\cite{Magnea:2018hab,Magnea:2018ebr}, the CoLoRFull
method
\cite{Somogyi:2005xz,Somogyi:2006cz,Somogyi:2006da,Somogyi:2006db,Somogyi:2008fc,Aglietti:2008fe,Somogyi:2009ri,Bolzoni:2009ye,Bolzoni:2010bt,DelDuca:2013kw,Somogyi:2013yk,DelDuca:2016ily}
and other approaches, e.g. the projection-to-Born method~\cite{Cacciari:2015jma}.

Despite the large number of available subtraction and slicing schemes,
it is fair to say that an optimal subtraction scheme, capable of
dealing with complex processes, is yet to be designed.
A set of criteria that should be considered when attempting the
construction of a subtraction scheme may include physical
transparency, scalability and locality as well as analyticity and
efficiency.
With these considerations in mind, the nested soft-collinear
subtraction scheme was introduced in Ref.~\cite{Caola:2017dug},
building on the sector-improved residue subtraction
scheme~\cite{Czakon:2010td,Czakon:2011ve,Czakon:2014oma}.
There it was shown that subtractions applied to gauge-invariant
scattering amplitudes, rather than to individual Feynman diagrams, can
be done in a nested fashion, yielding a somewhat simpler description.

As the name suggests, subtraction schemes handle infrared
singularities of real corrections by designing suitable subtraction
terms for soft and collinear divergences. Properly constructed
differences of real emission contributions and subtraction terms
become integrable over the full phase-space in four dimensions.
The subtraction terms, however, still need to be integrated in
$d=4-2\ep$ dimensions where soft and collinear singularities manifest
themselves as $1/\ep$ poles.

When the nested soft-collinear subtraction scheme is applied to
massless partons, there are two genuinely double-unresolved limits
that need to be addressed.
These are the double-soft limit, where two emitted partons become
soft, and the triple-collinear limit, where momenta of three partons
become collinear to each other.
For both of these cases, integrals over the phase space of unresolved
partons, subject to specific energy constraints dictated by the setup
of the nested soft-collinear subtraction scheme, have been
computed~\cite{Caola:2018pxp,Delto:2019asp}.
These results facilitate an analytic and fully-differential
description of colour-singlet production~\cite{Caola:2019nzf},
colour-singlet decay~\cite{Caola:2019pfz} and DIS-like processes
\cite{Asteriadis:2019dte}.
We note that these ``dipole-like'' building blocks should enable a
fully-differential NNLO QCD description of arbitrary processes.

The structure of IR singularities changes if massive quarks are
involved in a partonic process.
Indeed, since there are no collinear singularities related to massive
external legs, only soft singularities need to considered. They can be
subtracted using appropriate soft eikonal functions that have to be
integrated over the unresolved phase space.
The goal of this article is to start exploring the subtraction terms
that arise in NNLO QCD calculations for processes involving massive quarks
in the context of the nested soft-collinear subtraction scheme.
Specifically, we compute the integrated subtraction terms which are
required to describe double-soft emissions off two radiators of the
same mass in a back-to-back kinematics.

The remainder of this paper is organized as follows.
In Sec.~\ref{seq:prel}, we describe the nested
soft-collinear subtraction scheme and the single- and the
double-soft functions for massive radiators.
In Sec.~\ref{sec:phsp_int}, we integrate the single-soft and the
double-soft eikonal functions over the respective unresolved phase
space.
We discuss results in Sec.~\ref{sec:results} and conclude in
Sec.~\ref{sec:conc}

\section{Preliminary remarks}
\label{seq:prel}

In this section, we specify a physical setup, describe the idea
behind the nested soft-collinear subtraction scheme and
establish notations by writing down the factorisation formulae for
QCD amplitudes in the single-soft and the double-soft limits.
We conclude the section by defining sets of single- and double-soft 
emission integrals that need to be computed.

Our goal is to describe infrared (IR) singularities that arise in NNLO
QCD calculations for processes involving massive quarks.
In this work, we focus on the IR singularities that appear in the
double-real contribution to the decay of a colour-singlet particle $X$ to massive
quarks, i.e. a tree-level process
\begin{align}
  \label{eqn:def_part_proc}
  X
  \longrightarrow
  Q(p_A)
  +
  \bar{Q}(p_B)
  +
  f(k_1)
  +
  \bar{f}(k_2)\,.
\end{align}
In Eq.~\eqref{eqn:def_part_proc}, $Q$ stands for a massive quark while $f$ and
$\bar{f}$ denote a pair of massless partons (gluons or a
quark-antiquark pair).
Following Ref.~\cite{Caola:2017dug}, we write the contribution of
the partonic process in Eq.~\eqref{eqn:def_part_proc} to the decay
rate as
\begin{align}
  \label{eqn:def_sig_rr}
  \left\langle
  \DIFFL \Gamma_{\rm RR}
  \right\rangle
  ={}&
       \mathcal{N}
       \INT{}{}
       [\DIFFL k_1]
       [\DIFFL k_2]
       \theta(E_1-E_2)
       \DIFFL {\rm Lips}_{AB;12}
       | \mathcal{M} (A,B;1,2)|^2
       ~ \mathcal{F}(A,B;1,2)
        \nonumber \\
  ={}&
       \bigg\langle
       [\DIFFL k_1]
       [\DIFFL k_2]
       F_{LM}(A,B;1,2)
       \bigg\rangle  \,.
\end{align}
In Eq.~\eqref{eqn:def_sig_rr}, $\mathcal{N}$ includes normalization and symmetry factors,
$\DIFFL {\rm Lips}_{AB;12}$ denotes the Lorentz invariant phase-space
measure of the massive quark system, including the energy-momentum
conserving $\delta$-function, and $\mathcal{F}$ is the measurement
function of an arbitrary infrared-safe observable. All of these quantities
are then absorbed into the function $F_{LM}$ in Eq.~\eqref{eqn:def_sig_rr}.

Note that, following the original formulation of the nested
soft-collinear subtraction scheme, we have introduced an energy ordering for
the radiated partons $k_1$ and $k_2$, i.e. we require $E_1 > E_2$ in Eq.~\eqref{eqn:def_sig_rr}.
We define the phase-space element of a massless parton as
\begin{align}
  \label{eqn:PS_def}
  [\DIFFL k_i]
  ={}&
       \frac{ \DIFFFnopi{d-1}{{k_i}} }{2E_i} \theta(\Emax-E_i) \,.
\end{align}
In Eq.~\eqref{eqn:PS_def}, we introduced an energy cut-off $\Emax$
which is arbitrary but must be large enough so that it does not change
the value of the integral in Eq.~\eqref{eqn:def_sig_rr}, see
  Ref.~\cite{Caola:2017dug} for details.
The need for such a cut-off parameter will become clear later when the
double-soft limit of Eq.~\eqref{eqn:def_sig_rr} is discussed.  For
now, we only note that $\Emax$ breaks Lorentz invariance but
leaves rotational invariance intact.

\subsection{The nested soft-collinear subtraction scheme}
\label{subsec:nscss}

Infrared divergences of QCD amplitudes can be regulated by
introducing appropriate subtraction terms for all relevant kinematic configurations.
Within the nested soft-collinear subtraction scheme, such subtractions
are constructed in an iterative manner, starting from the double-soft limit.

The double-soft limit describes kinematic configurations where
energies of both emissions in Eq.~\eqref{eqn:def_sig_rr}, $E_1$ and $E_2$,
vanish at a comparable rate.
To describe this limit, we introduce a double-soft projection operator
$\DS$. For a generic amplitude $\mathcal{M}$ involving momenta
$k_1$ and $k_2$, we consider the scaling $E_1\sim E_2\sim \lambda$
and define 
\begin{align}
  \DS |\mathcal{M}(k_1,k_2)|^2
  ={}&
      \lim_{\lambda\to 0} \lambda^4 |\mathcal{M}(\lambda k_1, \lambda k_2)|^2 \,.
\end{align}
We then split the double-real contribution in Eq.~\eqref{eqn:def_sig_rr} into
double-soft regulated and unresolved parts, i.e.
\begin{align}
  \begin{aligned}
  \label{eqn:ds_subt}
  \bigg\langle [\DIFFL k_1] [\DIFFL k_2] F_{LM}(A,B;1,2) \bigg\rangle
  ={}&
       \bigg\langle [\DIFFL k_1] [\DIFFL k_2] ( I - \DS ) F_{LM}(A,B;1,2) \bigg\rangle
       \\
     &+
       \bigg\langle [\DIFFL k_1] [\DIFFL k_2] \DS F_{LM}(A,B;1,2) \bigg\rangle \,.
  \end{aligned}
\end{align}
The first term on the right-hand side is not divergent in the
double-soft limit. This term still contains singularities in the
single-soft limit, where $E_2\rightarrow0$, and in the collinear
limit, where the two emitted partons become collinear to each other.
Deriving a full subtraction would require us to remove these
singularities as well.  However, since integrated subtraction terms in
these two limits can be obtained in a rather straightforward manner
(see e.g. Ref.~\cite{Behring:2019oci}), in this paper we focus on the
second term on the right-hand side of Eq.~\eqref{eqn:ds_subt} and its
integration over the double-unresolved phase space. It reads
\begin{align}
\label{eqn:ds_ct}
\bigg\langle [\DIFFL k_1]& [\DIFFL k_2] \DS F_{LM}(A,B;1,2) \bigg\rangle \nonumber \\
={}& \mathcal{N} \INT{}{} [\DIFFL k_1]    [\DIFFL k_2]     \theta(E_1-E_2)  \DIFFL {\rm Lips}_{AB} 
 \, \DS | \mathcal{M} (A,B;1,2)|^2  ~ \mathcal{F}(A,B) \,.
\end{align}
We note that in the double-soft limit the momenta $k_1$ and $k_2$
completely decouple from the hard matrix element, from the
energy-momentum conserving $\delta$-function and from the measurement
function $\mathcal{F}$. This allows us to obtain integrals over the
double-unresolved phase space in a universal manner.
After the decoupling from the energy-momentum conservation, integrals
over $\DIFFL E_1$ and $\DIFFL E_2$ in Eq.~\eqref{eqn:ds_ct} are only limited by 
the cut-off parameter $\Emax$ introduced in Eq.~\eqref{eqn:PS_def}.

As a consequence of the factorization in the double-soft limit, the
reduced matrix element describes the Born-like process
\begin{align}
  X \longrightarrow Q(p_A) + \bar{Q}(p_B)
\end{align}
and the momenta $p_{A,B}$ are back-to-back in
the rest frame of the decaying particle $X$.
In this kinematic situation
\begin{align}
  \label{eqn:def_Xframe}
  p_A+p_B = p_{AB}  = E(1,+\beta \VEC{n}) + E(1,-\beta\VEC{n}) = ( 2E , \VEC{0} ) \ ,
\end{align}
and the heavy-quark momenta are on the mass shell
\begin{align}
  p_{A}^2 =p_{B}^2 = m^2\,.
\end{align}
We note that the the quark energy $E$ is half the mass of the decaying
particle, $E = M_X/2$, the vector $\VEC{n}$ describes the direction of
flight of the heavy quark in the rest frame of the decaying colour
singlet and
\begin{align}
  \beta
  ={}&
       \sqrt{1 - \frac{m^2}{E^2}}\,.
\end{align}
The threshold limit $E=m$ implies $\beta=0$.

\subsection{Eikonal functions for single- and double-soft emissions}
\label{sec:soft_fac}
Soft factorization formulas for generic QCD tree-level amplitudes
involving massless radiators and up to two soft partons were
studied, for example, in Ref.~\cite{Catani:1999ss}. This result
was extended to cover massive radiators in Ref.~\cite{Czakon:2011ve}
using the observation that eikonal currents are identical for massive
and massless emitters and that emitters' masses become relevant only
when eikonal currents are squared.

We begin with the single-gluon emission. The limit of an amplitude
that contains a gluon with a soft momentum $k$ reads
\begin{align}
  \label{eqn:fac_form_nlo_MES}
  \hat{S}_k | \mathcal{M}^g(\{p\},k) |^2
  =
  - g_{s,b}^{2} \SUM{i,j=1}{n} \mathcal{S}_{ij}(k)
  | \mathcal{M}^{(ij)}(\{p\}) |^2 \ ,
\end{align}
where $\{p\}=\{p_1,p_2,\ldots,p_n\}$ and the sum runs over all $n$ hard
emitters. The operator $\hat{S}_k$ extracts the
leading asymptotic behaviour of the matrix element in the soft limit,
$E_k\rightarrow0$. The single-eikonal
function $\mathcal{S}_{ij}(k)$ reads
\begin{align}
  \label{eqn:nlo_eik}
  \mathcal{S}_{ij}(k) = \frac{(p_i\cdot p_j)}{(p_i\cdot k)(p_j\cdot k)} \,.
\end{align}
The colour correlations in Eq.~\eqref{eqn:fac_form_nlo_MES} are
encoded in the reduced matrix element\footnote{To describe colour
  degrees of freedom we use colour-space notation from
  Ref.~\cite{Catani:1996vz}.}
\begin{align}
  | \mathcal{M}^{(ij)}(\{p\}) |^2
  = \langle \mathcal{M}(\{p\}) | \VEC{T}_i \cdot \VEC{T}_j | \mathcal{M}(\{p\}) \rangle \,.
\end{align}

The double-soft function that describes emission of two gluons with
momenta $k_1$ and $k_2$ reads
\begin{align}
  \label{eqn:eik_func_gg_gen}
  \DS
  | \mathcal{M}^{gg}(\{p\},k_1,k_2) |^2
  ={}
  g_{s,b}^{4}
  \bigg\{ &
            \frac{1}{2} \SUM{i,j,k,l=1}{n}
            \mathcal{S}_{ij}(k_1) \mathcal{S}_{kl}(k_2)
            | \mathcal{M}^{\{(ij),(kl)\}}(\{p\}) |^2
            \nonumber\\
          &
            - C_A \SUM{i,j=1}{n} \mathcal{S}_{ij}(k_1,k_2) | \mathcal{M}^{(ij)}(\{p\})|^2
            \bigg\} \ ,
\end{align}
where the additional colour correlated matrix element is defined as
\begin{align}
  \label{eqn:colour_corr_abel}
  | \mathcal{M}^{\{(ij),(kl)\}}(\{p\})|^2
  ={}
  \langle \mathcal{M}(p_1,\dots,p_n) |
  \{ \VEC{T}_i \cdot \VEC{T}_j ,\VEC{T}_k \cdot \VEC{T}_l\}
  | \mathcal{M}(p_1,\dots,p_n) \rangle \ ,
\end{align}
and the notation $\{\cdot,\cdot\}$ stands for an anticommutator in
colour space.
The first term on the right-hand side of Eq.~\eqref{eqn:eik_func_gg_gen} is the
\textit{abelian} contribution. It is simply a product of
single-eikonal factors defined in Eq.~\eqref{eqn:nlo_eik}.
Due to its factorized form, it is particularly easy to integrate this
term over the soft-gluons phase space.
The second, \textit{non-abelian} contribution is proportional to the
colour factor $C_A$.  It is given by the function
$\mathcal{S}_{ij}(k_1,k_2)$ which reads
\begin{align}
  \label{eqn:eik_func_gg_gen_nab}
  \mathcal{S}_{ij}(k_1,k_2)
  ={}&
       \mathcal{S}^{0}_{ij}(k_1,k_2) +
       \left[
       m_i^2 \mathcal{S}^{m}_{ij}(k_1,k_2)
       +
       m_j^2 \mathcal{S}^{m}_{ji}(k_1,k_2)
       \right] ,
\end{align}
where we note that the term in square brackets explicitly depends on
the squared masses of the emitters $m_{i}^2$ and $m_{j}^2$. Both
$\mathcal{S}^{0}_{ij}(k_1,k_2)$ and $\mathcal{S}^{m}_{ij}(k_1,k_2)$
implicitly depend on the masses.  The first term in
Eq.~\eqref{eqn:eik_func_gg_gen_nab}, $\mathcal{S}^{0}_{ij}(k_1,k_2)$, also
appears in the factorization formula for massless emitters
\cite{Catani:1999ss}; it reads
\begin{align}
    \label{eqn:ds_gg_nab_m0}
    \mathcal{S}^0_{ij}(k_1,k_2) &\nonumber\\
    ={}&   \frac{(1-\ep)}{(k_1\cdot k_2)^2} \frac{\left[(p_i\cdot k_1)(p_j\cdot k_2) + i\leftrightarrow j \right]}{(p_i\cdot k_{12})(p_j\cdot k_{12})} \nonumber \\
    & - \frac{(p_i\cdot p_j)^2}{2(p_i\cdot k_1)(p_j\cdot k_2)(p_i\cdot k_2)(p_j\cdot k_1)} \bigg[ 2 - \frac{\left[(p_i\cdot k_1)(p_j\cdot k_2) + i\leftrightarrow j \right]}{(p_i\cdot k_{12})(p_j\cdot k_{12})} \bigg] \nonumber\\
    & + \frac{(p_i\cdot p_j)}{2(k_1\cdot k_2) } \bigg[ \frac{2}{(p_i\cdot k_1)(p_j\cdot k_2)} + \frac{2}{(p_j\cdot k_1)(p_i\cdot k_2)} - \frac1{(p_i\cdot k_{12})(p_j\cdot k_{12})} \nonumber \\
    & ~~ \times \left( 4 + \frac{\left[(p_i\cdot k_1)(p_j\cdot k_2) + i\leftrightarrow j \right]^2}{(p_i\cdot k_1)(p_j\cdot k_2)(p_i\cdot k_2)(p_j\cdot k_1)} \right) \bigg] \ .
\end{align}
The other two contributions in Eq.~\eqref{eqn:eik_func_gg_gen_nab} are only
relevant for massive hard emitters. The function
$\mathcal{S}^{m}_{ij}(k_1,k_2)$ is given by~\cite{Czakon:2011ve}
\begin{align}
    \label{eqn:ds_gg_nab_m}
    \mathcal{S}^{m}_{ij}(k_1,k_2) &\nonumber\\
    ={}& - \frac1{4(k_1\cdot k_2)(p_i \cdot k_1)(p_i \cdot k_2)} + \frac{(p_i\cdot p_j) (p_j \cdot k_{12})}{2(p_i\cdot k_1)(p_j\cdot k_2)(p_i\cdot k_2)(p_j\cdot k_1)(p_i \cdot k_{12})} \nonumber\\
    & - \frac1{2(k_1\cdot k_2)(p_i\cdot k_{12})(p_j\cdot k_{12})} \left( \frac{(p_j\cdot k_1)^2}{(p_i\cdot k_1)(p_j\cdot k_2)} + \frac{(p_j\cdot k_2)^2}{(p_i\cdot k_2)(p_j\cdot k_1)} \right) \ .
\end{align}
Note that we use an abbreviation $k_{12}=k_1+k_2$ in
Eqs.~\eqref{eqn:ds_gg_nab_m0} and~\eqref{eqn:ds_gg_nab_m}.

When a soft quark-antiquark pair is emitted, the soft limit of the
matrix element is described by
\begin{align}
  \label{eqn:eik_func_qq_gen}
  \DS
  | \mathcal{M}^{q\bar{q}}(\{p\},k_1,k_2) |^2
  ={}
  g_{s,b}^{4} \ T_F \SUM{i,j=1}{n} \mathcal{I}_{ij}(k_1,k_2)
  | \mathcal{M}^{(ij)}(\{p\})|^2  \ ,
\end{align}
where $T_F = 1/2$. The soft function $\mathcal{I}_{ij}(k_1,k_2)$ is given by
\begin{align}
  \label{eqn:eik_func_qq_gen_I}
  \mathcal{I}_{ij}(k_1,k_2) = \frac{\left[ \left(p_i \cdot k_1 \right) \left(p_j \cdot k_2 \right) + i\leftrightarrow j \right] -\left(p_i \cdot p_j \right)\left(k_1 \cdot k_2 \right) }{\left(k_1 \cdot k_2 \right)^2\left(p_i \cdot k_{12} \right) \left( p_j \cdot k_{12} \right)} \ .
\end{align}

As we already mentioned, in the soft limit the dependence on the soft
gluon momenta drops out from the matrix element as well as from the
momentum conserving $\delta$-function.  For this reason, the eikonal
factors in Eqs.~\eqref{eqn:eik_func_gg_gen} and
\eqref{eqn:eik_func_qq_gen} can be integrated over the soft-gluons
phase space, irrespective of matrix elements that describe the
underlying hard process.

In the following, we explain how to do that in the case of two equal
mass emitters whose momenta $p_A$ and $p_B$ are back-to-back.
To simplify notations, we introduce the single-emission phase-space
integral
\begin{align}
  \label{eqn:int_ss_def}
  \mathcal{G}_{ij} ={}& \INT{}{} [\DIFFL k] \ \mathcal{S}_{ij}(k) \,,
\end{align}
where $ij \in \{AA, AB, BA, BB\}$ and the phase-space measure and
eikonal functions are defined in Eqs.~\eqref{eqn:PS_def}
and~\eqref{eqn:nlo_eik}, respectively.
For the double-emission phase-space integrals, we
distinguish between emissions of gluons and quarks and define
\begin{align}
  \begin{aligned}
    \label{eqn:int_ds_def}
    \mathcal{GG}_{ij} & = 
    \INT{E_2<E_1}{} [\DIFFL k_1] [\DIFFL k_2] \ \mathcal{S}_{ij}(k_1,k_2)
    \,, \\
    \mathcal{Q\bar{Q}}_{ij} &  = 
    \INT{E_2<E_1}{} [\DIFFL k_1] [\DIFFL k_2] \ \mathcal{I}_{ij}(k_1,k_2)
    \,,
  \end{aligned}
\end{align}
where, again, $ij \in \{AA, AB, BA, BB\}$.
We note that in case of the of back-to-back kinematics, integrated
subtraction terms $BB$ and $AA$, as well as $BA$ and $AB$, are equal
to each other. Therefore, in what remains, we will only consider cases
$ij=AA$ and $ij=AB$.

The integrals in Eqs.~\eqref{eqn:int_ss_def}
and~\eqref{eqn:int_ds_def} fully describe the integrated soft
subtraction terms in the decay process of Eq.~\eqref{eqn:def_part_proc}
and are an important ingredient for more complex processes,
such as heavy-quark pair production.
The computation of phase-space integrals in
Eq.~\eqref{eqn:int_ds_def} is the main goal of this paper.
We describe the details of the computation in the following section.
We note that a similar calculation was performed in
Ref.~\cite{Wang:2018vgu}, however, the unresolved phase space in that
paper was subject to a slightly different constraint.

\section{Phase-space integrals}
\label{sec:phsp_int}
In this section we present details of the calculation of the integrals
defined in Eqs.~\eqref{eqn:int_ss_def} and~\eqref{eqn:int_ds_def}.
We start with the single-soft emission to clarify notation and then
proceed to the double-emission case. The results of the latter
calculation are discussed in Section~\ref{sec:results}.

\subsection{Single-emission integrals}
\label{sec:ss_int}
We start with a brief discussion of single-emission integrals. The
first integral reads
\begin{align}
  \label{eqn:int_ss_ab}
    \G{AB}
    ={}&
    \INT{}{}
    [\DIFFL k] \
    \frac{ \scr{p_A}{p_B} }{ \scr{p_A}{k} \scr{p_B}{k} }
    \nonumber\\
    ={}&
    \frac{ (1+\beta^2)}{2}
    \INT{0}{\Emax}
    \frac{ \DIFFL E }{E^{1+2\ep}}
    \INT{}{} \frac{ \DIFFSPinopi{d-1}{k} }{ (1- \beta\VEC{n}\cdot\VEC{n}_k)(1 + \beta\VEC{n}\cdot\VEC{n}_k)  } \,,
\end{align}
where we have parametrised the gluon four momentum as $k =
E(1,\VEC{n}_k)$.  Further, we choose the reference frame in such a way
that the $z$-axis points in the $\VEC{n}$ direction. This yields
$(1\pm \beta\VEC{n}\cdot\VEC{n}_k) = (1\pm\beta\cos\theta)$ and, after
introducing $\eta = (1 - \cos \theta ) / 2$, we obtain
\begin{align}
  \label{eqn:int_ss_ab_apart}
  \G{AB}
  ={}&
       -\frac{(1+\beta^2)\Emax^{-2\ep}}{4\ep}
       \times\Svol{d-2}
       \FEYNINT{\eta}
       \left(
       \frac{\left[4\eta(1-\eta)\right]^{-\ep}}{\left[1-\beta(1-2\eta)\right]}
       +
       \frac{\left[4\eta(1-\eta)\right]^{-\ep}}{\left[1+\beta(1-2\eta)\right]}
       \right)
       \,,
\end{align}
where $\Svol{n} =  2\pi^{n/2} / \GF{n/2}$ denotes the volume of a unit sphere in $n$ dimensions.
The integral in Eq.~\eqref{eqn:int_ss_ab_apart} can be written as
a hypergeometric function of the type $\HYPGF{a}{b}{2b}{z}$,
which further simplifies to~\cite{abramowitz:1964}
\begin{align}
  \HYPGF{a}{b}{2b}{z}
  ={}&
       \left(1-z/2\right)^{-a} \, \HYPGF{a/2+1/2}{a/2}{b+1/2}{z^2/(2-z)^2} \,.
\end{align}
We find
\begin{align}
  \label{eqn:int_ss_ab_result}
  \G{AB}
  ={}&
       -\frac{ (1 +  \beta^2) \Emax^{-2\ep} }{4\ep}
       \times\Svol{d-1}
       \times \HYPGF{1}{1/2}{3/2-\ep}{\beta^2} \,.
\end{align}
The second integral, for a self-correlated emission, reads
\begin{align}
    \label{eqn:int_ss_aa_result}
    \G{AA}
    ={}&
    \INT{}{}
    [\DIFFL k] \
    \frac{m^2}{\scr{p_A}{k}^2} \nonumber\\
    ={}&
    - \frac{\Emax^{-2\ep}}{4\ep}
    \times \Svol{d-1}
    \times \left( 1- 2 \ep +2 \ep \HYPGF{1}{1/2}{3/2-\ep}{\beta^2} \right)\,.
\end{align}
Note that the hypergeometric function which appears in
Eqs.~\eqref{eqn:int_ss_ab_result} and~\eqref{eqn:int_ss_aa_result}
features an expansion in powers of $\ep$ in terms of classical
polylogarithms with arguments that involve square roots of $\beta$.
In order to simplify the expansion, we again rewrite the
hypergeometric function~\cite{abramowitz:1964} and find
\begin{align}
\label{eqn:2F1_bettervar}
    \HYPGF{1}{1/2}{3/2-\ep}{\beta^2}
    ={}&
    \frac{1-2\ep}{2\ep\beta}
    \left(\frac{2\beta}{1+\beta}\right)^{2\ep}
    \times\nonumber\\
    &\hspace{-3cm}
    \times
    \bigg\{
    \left(\frac{1-\beta}{1+\beta}\right)^{-\ep}
    \frac{\GF{1-2\ep}\GF{1+\ep}}{\GF{1-\ep}}
    -
    \HYPGF{\ep}{2\ep}{1+\ep}{\frac{1-\beta}{1+\beta}}
    \bigg\}
    \,.
\end{align}
Using \texttt{HypExp}~\cite{Huber:2007dx}, the hypergeometric function in Eq.~\eqref{eqn:2F1_bettervar} can be expanded as
\begin{align}
  &\hspace{-1cm}\HYPGF{\ep}{2\ep}{1+\ep}{\frac{1-\beta}{1+\beta}}\nonumber\\
  ={}&  1 + 2 \ep^2 \xLi{2}{\frac{1-\beta}{1+\beta}} + \ep^3 \bigg[ 4\zeta_3 + \frac{2\pi^2}{3} \xLn{\frac{2\beta}{1+\beta}}  -2 \xLn{\frac{1-\beta}{1+\beta}} \xLnp{2}{\frac{2\beta}{1+\beta}} \nonumber\\
  & - 4\xLn{\frac{2\beta}{1+\beta}} \xLi{2}{\frac{1-\beta}{1+\beta}} -2 \xLi{3}{\frac{1-\beta}{1+\beta}} - 4 \xLi{3}{\frac{2\beta}{1+\beta}} \bigg]  + \ORD{\ep^4}\,.
\end{align}
The results shown in Eqs.~\eqref{eqn:int_ss_ab_result} and
\eqref{eqn:int_ss_aa_result} were derived earlier in the
literature~\cite{Alioli:2010xd,Somogyi:2011ir}.

\subsection{Double-emission integrals}
\label{sec:ds_int}
We now turn to the calculation of the double-soft subtraction terms.
We need to compute the four functions $\mathcal{GG}_{AA}$,
$\mathcal{GG}_{AB}$, $\mathcal{Q\bar{Q}}_{AA}$ and
$\mathcal{Q\bar{Q}}_{AB}$ in Eq.~\eqref{eqn:int_ds_def}. To this end
we employ reverse unitarity~\cite{Anastasiou:2002yz} that has been
previously used for the computation of other integrated subtraction
terms~\cite{Caola:2018pxp,Delto:2019asp}.

\subsubsection*{Computational setup}
The integration measure for the two energy-ordered emissions in Eq.~\eqref{eqn:int_ds_def} reads
\begin{align}
  [\DIFFL k_1]
  [\DIFFL k_2] \Big|_{E_2 < E_1}
  ={}&
       \frac{ \DIFFFnopi{d-1}{{k_1}} }{2E_1}\,
       \frac{ \DIFFFnopi{d-1}{{k_2}} }{2E_2}\,
       \theta(\Emax-E_1) \,
       \theta(E_1-E_2)\,.
\end{align}
Note that all integrands, $\mathcal{S}_{ij}(k_1,k_2)$ and
$\mathcal{I}_{ij}(k_1,k_2)$, are homogeneous under uniform rescaling of
$E_1$ and $E_2$. For this reason we parametrise the energies as
\begin{align}
  E_1 ={}& \Emax \cdot x \,, &
  E_2 ={}& \Emax \cdot x \cdot z \,, &
\end{align}
and integrate over $x$ to obtain
\begin{align}
  \label{eqn:int_ds_gg_def_z}
  \GG{ij} ={}& - \frac{\Emax^{-4\ep}}{16\ep}  \FEYNINT{z} ~ z^{1-2\ep}  \INT{}{} \DIFFSPinopi{d-1}{12} ~ \mathcal{S}_{ij}(n_1,z \cdot n_2) \,, \\
  \label{eqn:int_ds_qq_def_z}
  \QQ{ij} ={}& - \frac{\Emax^{-4\ep}}{16\ep}  \FEYNINT{z} ~ z^{1-2\ep}  \INT{}{} \DIFFSPinopi{d-1}{12} ~ \mathcal{I}_{ij}(n_1,z \cdot n_2) \,,
\end{align}
where $n_i = (1,\VEC{n}_i)$, and the angular integration measure reads
$\DIFFSPinopi{d-1}{12} = \DIFFSPinopi{d-1}{1} \DIFFSPinopi{d-1}{2}$.

It remains to carry out angular and $z$ integrations in
Eqs.~\eqref{eqn:int_ds_gg_def_z} and~\eqref{eqn:int_ds_qq_def_z}.
However, the gluon emission case exhibits a strongly-ordered limit,
where the gluon with momentum $k_2$ is much softer than the gluon with
momentum $k_1$. Such behaviour results in a logarithmic divergence
in the $z$ integration at $z=0$, which prevents us from a naive
Taylor expansion of the integrand in $\ep$. The problem can be
ameliorated by using endpoint subtraction at $z=0$.  To accomplish
this, we extract the divergent part using the following formula
\begin{align}
  \mathcal{S}^{\text{s.o.}}_{ij}(n_1, n_2) = z^{-2} \lim_{z\to0} \left[ z^2 \mathcal{S}_{ij}(n_1,z \cdot n_2) \right] \,.
\end{align}
We note that it is beneficial to perform such a subtraction at the level of
the full integrand since the resulting expression fully accounts for gauge properties
of QCD amplitudes and, in variance to individual integrals, 
does not exhibit unphysical singularities.

Note that an emission of a soft quark-antiquark pair does not exhibit the
$z\rightarrow0$ singularity and, for this reason, does not require
additional subtraction.

To perform angular integrals in Eqs.~\eqref{eqn:int_ds_gg_def_z}
and~\eqref{eqn:int_ds_qq_def_z} we proceed as follows.  In the spirit
of reverse unitarity~\cite{Anastasiou:2002yz}, we rewrite
$\delta$-functions through cut propagators.  To this end, we first
rewrite the angular integration measures for both emissions as
\begin{align}
  \DIFFSPinopi{d-1}{i}
  ={}&
       4 \DIFFD{k_i}
       \DDP{k_i^2}
       \DD{\scr{k_i}{p_{AB}} - \xi_i \, p_{AB}^2/2}
       \left( p_{AB}^2/4 \right)^{\ep}
       \xi_i^{-1+2\ep} \,,
       \quad i=1,2\,,
\end{align}
with $\xi_1 = 1$ and $\xi_2 = z$. By applying Cutkosky
rules~\cite{Cutkosky:1960sp} backwards, we define cut loop integrals
\begin{align}
  \label{eqn:ang_int_cut_2l}
  \begin{aligned}
    \eGG{ij}(z,\beta,\ep)
    ={}&
    \INT{}{} \frac{ \DIFFD{k_1}  \DIFFD{k_2} ~~ \mathcal{S}_{ij} \left( k_1 , k_2 \right) }{[k_1^2]_c [k_2^2]_c  [k_1\cdot p_{AB}-2E^2]_c [ k_2\cdot p_{AB}-2E^2 z]_c }
    \,, \\
    \eGGso{ij}(z,\beta,\ep)
    ={}&
    \INT{}{} \frac{ \DIFFD{k_1}  \DIFFD{k_2} ~~ \mathcal{S}^{\text{s.o.}}_{ij} \left( k_1 , k_2 \right) }{[k_1^2]_c [k_2^2]_c  [k_1\cdot p_{AB}-2E^2]_c [ k_2\cdot p_{AB}-2E^2 z]_c }
    \,, \\
    \eQQ{ij}(z,\beta,\ep)
    ={}&
    \INT{}{} \frac{ \DIFFD{k_1}  \DIFFD{k_2} ~~ \mathcal{I}_{ij} \left( k_1 , k_2 \right) }{[k_1^2]_c [k_2^2]_c  [k_1\cdot p_{AB}-2E^2]_c [ k_2\cdot p_{AB}-2E^2 z]_c }
    \,.
  \end{aligned}
\end{align}
We note that the variable $z$ appears only in one of the cut
propagators and plays the role of an internal mass. We use the definitions
of Eq.~\eqref{eqn:ang_int_cut_2l} in Eqs.~\eqref{eqn:int_ds_gg_def_z}
and~\eqref{eqn:int_ds_qq_def_z} and write
\begin{align}
  \label{eqn:z_int_final_gg}
  \GG{ij}
  ={}&
       - \frac1{\ep}
       \left(\frac{\Emax}{E}\right)^{-4\ep}
       \bigg[ \FEYNINT{z} \left( \eGG{ij}(z,\beta,\ep)  - \eGGso{ij}(z,\beta,\ep)\right) \nonumber \\
       & \hphantom{-\frac{1}{\ep}\left(\frac{\Emax}{E}\right)^{-4\ep}\bigg[} +  \FEYNINT{z} ~\eGGso{ij}(z,\beta,\ep) \bigg]
       \,, \\
  \label{eqn:z_int_final_qqb}
  \QQ{ij}
  ={}&
       - \frac1{\ep}
       \left(\frac{\Emax}{E}\right)^{-4\ep}
       \FEYNINT{z} ~ \eQQ{ij}(z,\beta,\ep)
       \,.
\end{align}
After mapping angular integrals onto ordinary loop integrals with cut
propagators, we employ standard techniques of loop calculations to
compute the integrals that appear in Eqs.~\eqref{eqn:z_int_final_gg}
and~\eqref{eqn:z_int_final_qqb}.

\subsubsection*{IBP reduction}
We apply integration-by-parts (IBP) techniques~\cite{Chetyrkin:1981qh}
to the integrands of Eqs.~\eqref{eqn:z_int_final_gg}
and~\eqref{eqn:z_int_final_qqb} to express them in terms of a few
master integrals.  The integrands consist of two-loop cut integrals
\begin{align}
\label{eqn:def_topos}
  T^{a_1,a_2,a_3}(\alpha_1,\alpha_2,\alpha_3)
  ={}&
       \left(E^2\right)^{-d+4+\SUM{i=1}{3}\alpha_i}
       \INT{}{}
       \frac{
       \DIFFD{k_1}  \DIFFD{k_2}
       }{
       \Dcut\,
       D_{a_1}^{\alpha_1}\,
       D_{a_2}^{\alpha_2}\,
       D_{a_3}^{\alpha_3}
       }
       \equiv{}
       \Imast{
         \PROD{i=1}{3} \frac{1}{D_{a_i}^{\alpha_i}}
       } \,,
\end{align}
where the propagators to be cut are given by
\begin{align}
  \Dcut
  ={}&
       [k_1^2]_c \,
       [k_2^2]_c \,
       [k_1\cdot p_{AB}-2E^2]_c \,
       [ k_2\cdot p_{AB}-2E^2 z]_c\,,
\end{align}
and the three ordinary propagators $D_{a_i}$ per topology
$T^{a_1,a_2,a_3}$ are drawn from a set
\begin{align}
  \label{eqn:uncut_props}
  D_{1,\dots,7}
  ={}&
       \left\{
       \scr{p_A}{k_1},\,
       \scr{p_B}{k_1},\,
       \scr{p_A}{k_2},\,
       \scr{p_B}{k_2},\,
       \scr{k_1}{k_2},\,
       \scr{p_A}{k_{12}},\,
       \scr{p_B}{k_{12}}
       \right\}\,.
\end{align}
The variables $\alpha_i$ in Eq.~\eqref{eqn:def_topos} refer to powers
of propagators in integrals in a certain topology $T^{a_1,a_2,a_3}$.
The prefactor in Eq.~\eqref{eqn:def_topos} was chosen to render
integrals dimensionless.  To express all integrals in
Eq.~\eqref{eqn:ang_int_cut_2l} through these topologies, we use the
following list of linear relations between propagators
\begin{align}
  \begin{aligned}
    D_1 + D_3 ={}& D_6\,, ~~ & D_2 + D_4 &  = D_7\,, \\
    D_1 + D_2 ={}& 2 E^2\,,~~ &   D_3 + D_4 & = 2 E^2 \, z\,,
  \end{aligned}
\end{align}
where the last two equations follow from the cut constraints.

We use \texttt{Reduze2}~\cite{vonManteuffel:2012np} to express
integrals shown in Eq.~\eqref{eqn:ang_int_cut_2l} through master integrals. We write
\begin{align}
  \eXX{ij}(z,\beta,\ep)
  ={}&
       \VEC{R}^{\mathcal{XX}}_{ij}(z,\beta,\ep)
       \cdot
       \VEC{I}(z,\beta,\ep)
       \,,
\end{align}
where $\VEC{R}^{\mathcal{XX}}_{ij}(z,\beta,\ep)$ are vectors of
reduction coefficients and $\VEC{I}(z,\beta,\ep)$ stands for a
vector constructed out of thirteen master integrals
grouped into five topologies. The first integral is the
phase-space volume
\begin{align}
  I_1
  ={}
  \Imast{1}
  ={}&
       z^{1-2\ep} \,
       \frac{ \left(\Svol{d-1}\right)^2 }{16}
       \,,
       \label{eqn:phsp-volume}
\end{align}
and the remaining twelve integrals are given by
\begin{align}
  \label{eqn:masters_lap}
  \begin{aligned}
  I_{2,\ldots,4} & = \left\{ \Imast{\frac1{D_3}}, \Imast{\frac{1}{D_2 D_3}}, \Imast{\frac{1}{D_2 D_3 D_5}} \right\} \ \subset T^{2,3,5}  \,, \\
  I_{5,\ldots,9} & = \left\{ \Imast{\frac{D_2}{D_6}}, \Imast{\frac{D_5}{D_6}}, \Imast{\frac{1}{D_6}}, \Imast{\frac{1}{D_2 D_6}}, \Imast{\frac{1}{D_2 D_5 D_6}} \right\} \ \subset T^{2,5,6}  \,, \\
  I_{10}         & = \left\{ \Imast{\frac{1}{D_2 D_7}} \right\} \ \subset T^{2,5,7}\,, \\
  I_{11,12}      & = \left\{ \Imast{\frac{1}{D_4 D_6}}, \Imast{\frac{1}{D_4 D_5 D_6}} \right\}\ \subset T^{4,5,6} \,, \\
  I_{13}         & = \left\{ \Imast{\frac{1}{D_4 D_7}} \right\}\ \subset T^{4,5,7}\,.
  \end{aligned}
\end{align}
We note that the gluon emission contribution $\eGG{ij}(z,\beta,\ep)$
requires the full set of master integrals. On the other hand, the
strongly ordered contribution $\eGGso{ij}(z,\beta,\ep)$ requires
master integrals $I_{1,\dots,4}$ and the quark-emission contribution
$\eQQ{ij}(z,\beta,\ep)$ only depends on integrals $I_{1,5,\dots,7}$.

\subsubsection*{Differential equations}
Having obtained a set of master integrals we employ the method of
differential
equations~\cite{Kotikov:1990kg,Remiddi:1997ny,Gehrmann:1999as} to
compute them.
To this end, we derive a closed system of first order partial differential
equations for the master integrals $\VEC{I}$ as functions of variables
$\beta$ and $z$. We then cast the differential equations into the
$\ep$-homogeneous form~\cite{Henn:2013pwa} by changing the basis of master
integrals
\begin{align}
  \VEC{I}
  ={}&
       \Tcan \VEC{J}
       \,.
\end{align}
Here $\Tcan$ is the transformation that brings master integrals into
their so-called canonical basis $\VEC{J}$.
In general, finding a canonical basis or, equivalently, constructing a
transformation $\Tcan$ is a complicated task. In our case, we accomplish
this by using the algorithmic approach suitable for
multi-scale problems proposed in Ref.~\cite{Meyer:2016slj} and
implemented in the \texttt{CANONICA} package~\cite{Meyer:2017joq} for
\texttt{Mathematica}.
This transformation can also found using the approach of
Ref.~\cite{Lee:2014ioa} implemented in a private \texttt{Mathematica}
tool \texttt{Libra}.\footnote{We wish to thank Roman Lee for giving us
  access to the \texttt{Libra} package.} In this case, a sequential
application of the algorithm of Ref.~\cite{Lee:2014ioa} is required.

In the canonical basis $\VEC{J}$, differential equations take the
$\ep$-homogeneous form
\begin{align}
  \label{eqn:deq_can}
  \partial_x \VEC{J}
  ={}&
       \ep \,
       \hat{M}_x \,
       \VEC{J} \,,
\end{align}
with $x \in \{ z, \beta \}$.
The matrices $\hat{M}_z$ and $\hat{M}_\beta$ feature simple poles
and can be written as
\begin{align}
  \hat{M}_x
  ={}&
       \SUM{x_i \in \mathcal{A}_x }{}
       \frac{ \hat{m}_{x_i} }{ x - x_i }
       \,.
       \label{eqn:mat-residues}
\end{align}
In Eq.~\eqref{eqn:mat-residues}, the residue matrices $\hat{m}_{x_i}$
are composed of rational numbers and the poles $x_i$ are drawn from
the two alphabets
\begin{align}
  \mathcal{A}_z
  ={}&
       \left\{
       0 ,\,
       -1 ,\,
       \frac{-2}{1\pm \beta} ,\,
       -\frac{(1\pm \beta)}{2} ,\,
       -\frac{1-\beta}{1+\beta} ,\,
       -\frac{1+\beta}{1-\beta} \,
       \right\}
       \label{eqn:z-alphabet}\,,
  \\
  \mathcal{A}_\beta
  ={}&
       \left\{
       0 ,\,
       \pm 1 ,\,
       \pm(1+2z) ,\,
       \pm \frac{1+z}{1-z} ,\,
       \pm \frac{2+z}{z} \,
       \right\}\,.
       \label{eqn:b-alphabet}
\end{align}

Thanks to the $\ep$-homogeneous form of the differential equations in
Eq.~\eqref{eqn:deq_can}, the $\ep$ expansion of the functions
$\VEC{J}(z,\beta)$ can be obtained by recursive integration of the
right-hand side. Since matrices $\hat{M}_x$ contain only simple poles,
the result can be expressed in terms of linear combinations of
Goncharov Polylogarithms (GPLs)~\cite{Goncharov:1994} that depend on
$z$ and $\beta$ and constants of integration.
Note that, since we are interested in a final integration over the
variable $z$ in Eqs.~\eqref{eqn:z_int_final_gg}
and~\eqref{eqn:z_int_final_qqb}, it is beneficial to write master
integrals in such a way that $z$ appears {\it only} as an argument of
the GPLs.
For this reason, at each order in $\ep$, we first integrate the system
of differential equations with respect to $z$.  A constant of
integration in this case is an unspecified function of $\beta$.
To determine this function, we substitute the solution into the
differential equations in $\beta$, and explicitly check that the
resulting differential equations are $z$ independent.  After
integration over $\beta$, all master integrals are expressed in terms
of GPLs, $\GPL{\vec{z}_0}{z}$ and $\GPL{\vec{\beta}_0}{\beta}$, where the
elements of $\vec{z}_0$ are drawn from the alphabet $\mathcal{A}_z$,
cf. Eq.~\eqref{eqn:z-alphabet}, and elements in $\vec{\beta}_0$
belong to the $z$-independent part of the alphabet
$\mathcal{A}_{\beta}$ in Eq.~\eqref{eqn:b-alphabet},
i.e. $\tilde{\mathcal{A}}_{\beta}= \{0, -1, +1\}$.

This concludes the computation of master integrals up to constants of
integration.  These constants are determined by calculating suitable
boundary conditions as we discuss in the next section.

\subsubsection*{Boundary conditions}
We find it suitable to determine constants of integration by
computing master integrals in the threshold limit $\beta \to 0$.  This
limit is particularly convenient, since many of the
integrals simplify. This happens because in that limit the
dependencies of all scalar products on quark momenta disappear. For
example
\begin{align}
  p_A\cdot(k_{1}+k_{2})
  ={}&
  E^2[ (1+z) - \beta \VEC{n}(\VEC{n}_1+z\VEC{n}_2) ]
  ~~\xrightarrow{\beta \to 0}{}~~
  E^2(1+z)\,.
\end{align}

By inspecting master integrals in Eq.~\eqref{eqn:masters_lap}, we
observe that
\begin{align}
  \lim_{\beta \to 0}
  \VEC{I}(z,\beta,\ep)
  ={}&
  \VEC{F}(z,\ep)
  + \mathcal{O}(\beta)\,.
\end{align}
Moreover, we find that all entries, except for the first diagonal
element of the canonical transformation matrix $\Tcan$ are suppressed
as $\mathcal{O}(\beta)$ and therefore vanish in the threshold limit.
The transformation matrix in the threshold limit reads
\begin{align}
  \lim_{\beta \to 0}\,
  \Tcan
  ={}&
  \begin{pmatrix}
    & \sfrac{1}{z}  & 0      & & \cdots \\
    & 0      & 0      & & \cdots \\
    & \vdots & \vdots & & \ddots \\
  \end{pmatrix}\,.
\end{align}
This means that to fix all integration constant we only need the
phase-space master integral $I_1$, which is straightforward to compute,
cf. Eq.~\eqref{eqn:phsp-volume}.

After fixing all the integration constants using boundary conditions, we
transform master integrals $\VEC{J}$ into the original basis
$\VEC{I}$.
We check the resulting expressions numerically for several values
of $\beta$ and $z$.

\subsubsection*{Integration over $z$}
\label{sec:z-int}
Having computed the required master integrals, we obtain the
integrands in Eqs.~\eqref{eqn:z_int_final_gg}
and~\eqref{eqn:z_int_final_qqb} and perform the integration over
$z$. The masters integrals $\VEC{I}$ of Eqs.~\eqref{eqn:phsp-volume}
and~\eqref{eqn:masters_lap} allow us to express the functions
$\eXX{ij}(z,\beta,\ep)$ in terms of rational functions of $z$, $\beta$
and GPLs of $z$ and $\beta$ with $z$-independent letters.
Such a representation enables the final $z$ integration in
Eqs.~\eqref{eqn:z_int_final_gg} and~\eqref{eqn:z_int_final_qqb} in a
straightforward manner.
We note that after obtaining the primitive, the $z \to 0$ limit
features spurious $1/z^n$ poles and needs to be taken with care.
We use \texttt{PolyLogTools}~\cite{Duhr:2019tlz} to expand all GPLs
around $z=0$ up to the order required to cancel these $1/z^n$ poles
and facilitate $z$-integration over the interval 
$0 < z < 1$. 
We report results for the  functions $\mathcal{GG}_{AA}$,
$\mathcal{GG}_{AB}$, $\mathcal{Q\bar{Q}}_{AA}$ and
$\mathcal{Q\bar{Q}}_{AB}$  in the next section.

\section{Results}
\label{sec:results}

In this section, we present some results for the integrated
double-soft subtraction terms, cf. Eq.~\eqref{eqn:int_ds_def}. We write
\begin{align}
\label{eqn:results-gg-qq}
\begin{aligned}
  \GG{ij}
  ={}&
  \frac{ \Emax^{-4\ep} }{16}
  \left(\Svol{d-1}\right)^2
  \times \ggx{ij}(\beta,\ep)\,,
  \\
  \QQ{ij}
  ={}&
  \frac{ \Emax^{-4\ep} }{16}
  \left(\Svol{d-1}\right)^2
  \times \qqx{ij}(\beta,\ep)\,,
\end{aligned}
\end{align}
with $\Svol{n}$ defined after Eq.~\eqref{eqn:int_ss_ab_apart}.
Four results for functions $\ffx{ij}(\beta,\ep)$ can be found
in an ancillary file provided with this submission.
They are expressed through GPLs of $\beta$ up to weight four, with
integer letters drawn from the alphabet
\begin{align}
  \mathcal{A}=\{ 0 , \pm1 , \pm 3  \} \,.
\end{align}
We use a private implementation of the \textit{super-shuffle}
identities described in Ref.~\cite{Frellesvig:2018lmm} to translate
the expressions obtained from the integration over $z$,
cf. \SEC{ds_int}, into such a \textit{fibration} basis. We note that
all GPLs appearing in Eq.~\eqref{eqn:results-gg-qq} are manifestly
real in the physical region $\beta\in[0,1]$.  For a numerical
evaluation of GPLs one can resort to publicly available
programs~\cite{Vollinga:2004sn,Naterop:2019xaf}. The functions
$\ffx{ij}(\beta,\ep)$ were checked numerically using an
adaptation of the numerical routine from
Ref.~\cite{Behring:2019oci}.

While functions $\ggx{AA}$ and $\ggx{AB}$, that describe the emission
of two soft gluons, feature $1/\ep^3$ poles, the functions $\qqx{AA}$
and $\qqx{AB}$, related to quark pair emissions, start only at
$1/\ep^2$.
This happens because the latter case does not exhibit a strongly
ordered soft divergence, cf. Eq.~\eqref{eqn:z_int_final_qqb}.  The
expressions for $1/\ep$ poles of the functions $\ffx{ij}(\beta,\ep)$
consist only of harmonic polylogarithms (HPLs)~\cite{Remiddi:1999ew} of $\beta$ up to weight
three.  We rewrite them in terms of independent classical
polylogarithms~\cite{Duhr:2011zq} and find
\begin{align}
  \ggx{AA}(\beta,\ep)
  ={}&
  -\frac1{8\ep^3}
  +
  \frac1{\ep^2} \frac1{4\beta}
  \bigg\{
  \Ln{x_\beta} + \beta
  \bigg\}
  +
  \frac1{\ep}
  \frac1{4\beta}
  \bigg\{
  2 \beta -3 \Ln{x_\beta} -8\beta\Ln{2}\nonumber \\
  & -2 \left[ \Li{2}{y^-_\beta} + \Li{2}{\beta} -\Li{2}{-\beta} \right] + {y^-_\beta}\Lnp{2}{x_\beta} \nonumber \\
  & - \Lnp{2}{y^-_\beta}+ \zeta_2
  \bigg\}
  +\ORD{\ep^0} \,,
  \\
  \ggx{AB}(\beta,\ep)
  ={}&
  \frac1{\ep^3}
  \frac1{8\beta}
  \bigg\{
  3 \beta + 2 z_\beta\Ln{x_\beta}
  \bigg\}
  - \frac1{\ep^2}
  \frac1{24\beta^2}
  \bigg\{
  32\beta^2 + \beta\left(31+13\beta^2 \right)\Ln{x_\beta} \nonumber \\
  & + 12 z_\beta\beta \left[ \Li{2}{\beta}-\Li{2}{-\beta} \right] + 3z_\beta^2 \Lnp{2}{x_\beta}
  \bigg\} \nonumber \\
  & - \frac1{\ep}\frac1{72\beta^2}
  \bigg\{ 104 \beta^2 + 27z_\beta^2\zeta_3 -120\beta^2\Ln{2} \nonumber \\
  & + 36 z_\beta^2 \left( \Li{3}{x_\beta}-\Li{3}{y^-_\beta}-\Li{3}{y^+_\beta} \right) + 72 \beta z_\beta \left( \Li{3}{\beta}-\Li{3}{-\beta} \right) \nonumber \\
  & + 2\beta\left(62\beta^2-25\right)\Ln{x_\beta} - 12 \beta\left(4\beta^2+13\right)\left(\Li{2}{\beta}-\Li{2}{-\beta}\right) \nonumber \\
  & + 6\beta\left(\beta^2-2\right)\left(\zeta_2-2\Li{2}{y^-_\beta} -\Lnp{2}{y^-_\beta} \right) \nonumber \\
  &  -18z_\beta^2\Ln{x_\beta}\left(\Li{2}{\beta}-\Li{2}{-\beta}\right)  \nonumber \\
  & -3 \left( 24+2\beta+9\beta^2-\beta^3+12\beta^4 \right)\Lnp{2}{x_\beta} \nonumber \\
  & -132 \beta z_\beta \Ln{2}\Ln{x_\beta} -18\zeta_2 z_\beta^2\left(3\Ln{x_\beta}-2\Ln{y^-_\beta}\right) \nonumber \\
  & + 18 z_\beta^2 \Ln{\beta}\Lnp{2}{x_\beta} +6 z_\beta \left(3+2\beta+3\beta^2\right)\Lnp{3}{x_\beta} \nonumber \\
  & + 6 z_\beta^2 \left( 3 \Ln{x_\beta}\Lnp{2}{y^-_\beta} -2 \Lnp{3}{y^-_\beta} -6 \Lnp{2}{x_\beta}\Ln{y^-_\beta} \right)
  \bigg\} +\ORD{\ep^0} \,,
  \\
  \qqx{AA}(\beta,\ep)
  ={}&
  -\frac1{4\ep^2}
  + \frac1{\ep}
  \frac1{4\beta}
  \bigg\{
  6\beta -4\beta\Ln{2} +\Ln{x_\beta}
  \bigg\}
  +\ORD{\ep^0} \,,
  \\
  \qqx{AB}(\beta,\ep)
  ={}&
  \frac1{\ep^2}\frac1{12\beta}
  \bigg\{
  z_\beta \Ln{x_\beta} -\beta
  \bigg\}
  + \frac1{\ep}\frac1{72\beta}
  \bigg\{
  34\beta - \left(37\beta^2+43\right) \Ln{x_\beta} \nonumber \\
  & -24 z_\beta \left( \Li{2}{y^-_\beta} + \Li{2}{\beta} - \Li{2}{-\beta}  \right) -24 \beta \Ln{2} \nonumber \\
  & + 6 z_\beta \left( \Lnp{2}{x_\beta} - 2 \Lnp{2}{y^-_\beta} +4 \Ln{2}\Ln{x_\beta} + 2\zeta_2 \right)
  \bigg\}
  +\ORD{\ep^0} \,,
\end{align}
where we used the abbreviations
\begin{align}
  x_\beta ={}& \frac{1-\beta}{1+\beta} \,, &
  y^{\pm}_\beta ={}& \frac{1\pm\beta}{2}\,, &
  z_\beta ={}& 1+\beta^2 \,.&
\end{align}

Even though expressions for the finite parts of the functions
$\ffx{ij}(\beta,\ep)$ are rather long, they simplify in certain
limits. In what follows, we present the expansions in the threshold
limit, $\beta \to 0$, and the high-energy limit, $\beta \to 1$.

We begin with the threshold limit, where the energies of the emitting
quarks are close to their masses, i.e.  $E \approx m$, which implies
$\beta \ll 1$.
We perform a Taylor expansion in small $\beta$ and find
\begin{align}
  \label{eqn:results-f-threshold}
  \ggx{AA}(\beta\approx0,\ep)
  ={}&
  -\frac1{8\ep^3}-\frac1{4\ep^2}+\frac{1-2\Ln{2}}{\ep} +2 \left( 2 \Ln{2}-1-\frac{\pi^2}{6}\right)
  \nonumber \\
  & + \beta^2\left[ - \frac1{6\ep^2} - \frac{4}{9\ep} + \left( \frac1{27} - \frac{8}{3} \Ln{2} \right) \right] +\ORD{\beta^4} \,,
  \\
  \ggx{AB}(\beta\approx0,\ep)
  ={}&
  -\frac1{8\ep^3}-\frac1{4\ep^2}+\frac{1-2\Ln{2}}{\ep} +2 \left( 2 \Ln{2}-1-\frac{\pi^2}{6}\right) \nonumber \\
  & + \beta^2\bigg[ -\frac{2}{3\ep^3} - \frac1{2\ep^2} + \frac1{\ep}\left(1-\frac{44}{9} \Ln{2} \right) \nonumber \\
  & + \left(\frac{104}{27} \Ln{2} -\frac1{3} -\frac{22}{27}\pi^2 \right) \bigg] +\ORD{\beta^4} \,,
  \\
  \qqx{AA}(\beta\approx0,\ep)
  ={}&
  -\frac1{4\ep^2}+\frac{1-\Ln{2}}{\ep}+\left(4 \Ln{2}-\frac{3}{2}-\frac{\pi^2}{6}\right) \nonumber \\
  & + \beta^2 \left[ -\frac1{6\ep} + \left( \frac{13}{18} - \frac{4}{3}\Ln{2} \right)  \right] +\ORD{\beta^4} \,,
  \\
  \qqx{AB}(\beta\approx0,\ep)
  ={}&
  -\frac1{4\ep^2}+\frac{1-\Ln{2}}{\ep}+\left(4 \Ln{2}-\frac{3}{2}-\frac{\pi^2}{6}\right) \nonumber \\
  &+ \beta^2\left[ -\frac{2}{9\ep^2} +\frac1{\ep}\left(\frac{25}{54} - \frac{8}{9}\Ln{2} \right) +\left( \frac{23}{162} - \frac{4}{27} \pi^2 + \frac{44}{27} \Ln{2} \right) \right] \nonumber \\
  & +\ORD{\beta^4}\,.
\end{align}
Note that the leading terms in Eq.~\eqref{eqn:results-f-threshold} are equal for emitters
in a back-to-back kinematics ($AB$) and self-correlated emissions ($AA$), i.e.
\begin{align}
\ffx{AA}(\beta,\ep) = \ffx{AB}(\beta,\ep) + \ORD{\beta^2}\,.
\end{align}
This is the case, since in the threshold limit, $\beta=0$, the spatial parts of momenta
$p_A$ and $p_B$ vanish.

In the high-energy limit, the energies of the emitting quarks
are much larger than their masses, $E \gg m$, which implies
$\beta \approx 1$.
Expanding in $(1-\beta)$, we find
\begin{align}
\label{eqn:results-f-highE}
  \ggx{AA}(\beta\approx1,\ep)
  ={}&
  -\frac1{8\ep^3}+\frac{1-\Ln{2}}{4\ep^2}+\frac1{2\ep}\left(1-\frac{\pi^2}{6}-\frac{5}{2}\Ln{2}-\frac1{2}\Lnp{2}{2}\right) \nonumber \\
  &+ \left( \frac{21}{2}\Ln{2} -3 -\frac{\pi^2}{6}\Ln{2} -\frac{\pi^2}{24} -\frac1{6}\Lnp{3}{2} - \frac{7}{4} \Lnp{2}{2} -\frac{\zeta_3}{2} \right)\nonumber \\ 
  & + \Ln{1-\beta} \bigg[ \frac1{4\ep^2} + \frac1{\ep} \left(\frac1{2}\Ln{2} -\frac{3}{4} \right) +\left( \frac{\pi^2}{6}-\frac1{2} + 3 \Ln{2} + \frac1{2}\Lnp{2}{2} \right) \bigg] \nonumber \\
  & - \Lnp{2}{1-\beta} \left[ \frac1{4\ep} + \left( \frac1{2}\Ln{2} - \frac{3}{4} \right) \right] + \frac1{6} \Lnp{3}{1-\beta} + \ORD{1-\beta}\,, \\
  \ggx{AB}(\beta\approx1,\ep) 
  ={}& \frac1{\ep^3}\left( \frac{3}{8}-\frac1{2}\Ln{2} \right) + \frac1{\ep^2}\left(\frac{11}{6} \Ln{2} -\frac{4}{3}-\frac{\pi ^2}{4}-\frac1{2}\Lnp{2}{2} \right) \nonumber \\
  & + \frac1{\ep} \left( \frac{13 \pi ^2}{18} -3 \zeta_3-\frac{13}{9}-\frac1{3}\Lnp{3}{2}-\frac{11}{6}\Lnp{2}{2}+\frac{97}{36}\Ln{2}-\frac{5\pi ^2}{12}\Ln{2} \right) \nonumber \\
  & + \bigg( 6\Li{4}{\frac1{2}}+\frac{7 \zeta_3 }{3}+\frac{5\zeta_3}{2} \Ln{2} +\frac{1787}{108}+\frac{179 \pi ^2}{108}-\frac{13 \pi ^4}{48}+\frac1{12}\Lnp{4}{2} \nonumber \\
  & + \frac{11}{9}\Lnp{3}{2}+\frac{881}{36}\Lnp{2}{2}-\frac{2\pi ^2}{3}\Lnp{2}{2}-\frac{2059}{54}\Ln{2}-\frac{13\pi ^2}{18} \Ln{2} \bigg) \nonumber \\
  & + \Ln{1-\beta} \bigg[ \frac1{2\ep^3} + \frac1{\ep^2} \left(\Ln{2}-\frac{11}{6} \right) + \frac1{\ep} \left( \Lnp{2}{2} + \frac{5\pi^2}{12} - \frac{37}{36}  \right) \nonumber \\
  & + \left( \frac{11\zeta_3}{4}+\frac{491}{27}-\frac{10 \pi ^2}{9}+\frac{2}{3}\Lnp{3}{2}-\frac{163}{6}\Ln{2}+\frac{5\pi ^2}{6} \Ln{2}  \right) \bigg] \nonumber \\
  & + \Lnp{2}{1-\beta} \bigg[ -\frac1{2\ep^2} -\frac1{\ep} \left(\Ln{2}-\frac{11}{6} \right) - \left( \Lnp{2}{2} + \frac{5\pi^2}{12} - \frac{37}{36}  \right) \bigg] \nonumber \\
  & + \Lnp{3}{1-\beta} \left[ \frac1{3\ep} +\left( \frac{2}{3} \Ln{2} -\frac{11}{9} \right) \right] - \frac1{6} \Lnp{4}{1-\beta} + \ORD{1-\beta} \,,
  \\
  \qqx{AA}(\beta\approx1,\ep)
  ={}&
  -\frac1{4\ep^2}+\frac1{\ep}\left(\frac{3}{2}-\frac{5}{4}\Ln{2}\right)+\left(\frac{43}{4}\Ln{2}-\frac{7}{4}\Lnp{2}{2}-6-\frac{5\pi^2}{24}\right) \nonumber \\
  & + \Ln{1-\beta} \left[ \frac1{4\ep} +\left( 3\Ln{2}-\frac{11}{4} \right) \right] - \frac1{4}\Lnp{2}{1-\beta} +\ORD{1-\beta} \,,
  \\
  \qqx{AB}(\beta\approx1,\ep)
  ={}&
  -\frac1{\ep^2}\left(\frac{1}{12}+\frac1{6}\Ln{2}\right) +\frac1{\ep}\left(\frac{17}{36}-\frac{\pi ^2}{9}-\frac{5}{6}\Lnp{2}{2}+\frac{7}{9}\Ln{2}\right) \nonumber \\
  & +\left(\frac{77 \pi ^2}{108} -\frac{13 \zeta_3}{6}-\frac{161}{54}-\frac1{9}\Lnp{3}{2}+\frac{44}{9}\Lnp{2}{2}+\frac{31}{27}\Ln{2}-\frac{5 \pi ^2}{9} \Ln{2}  \right)\nonumber \\
  & + \Ln{1-\beta} \left[ \frac1{6\ep^2} +\frac1{\ep} \left(\Ln{2}-\frac{10}{9}\right) +\left(\frac{139}{54}+\frac{2 \pi ^2}{9}+\Lnp{2}{2}-\frac{17}{3}\Ln{2}\right) \right] \nonumber \\
  &-\Lnp{2}{1-\beta} \left[\frac1{6\ep} + \left(\Ln{2}-\frac{10}{9} \right) \right] + \frac1{9} \Lnp{3}{1-\beta} +\ORD{1-\beta}  \,.
\end{align}
Note that these expressions contain logarithms of the form
$\Lnp{n}{1-\beta}$, which are divergent in the $\beta \to 1$ limit.
These logarithms are related to quasi-collinear divergences that
appear once the mass of the emitter, which screens the actual
collinear divergences, becomes small compared to the overall
energy. In the massless calculation~\cite{Caola:2018pxp}, all
$\Ln{1-\beta}$ terms manifest themselves as additional poles in
$1/\ep$.

\section{Conclusions}
\label{sec:conc}
In this paper, we presented analytic results for the integrated double-soft
subtraction terms that are needed in the context of the
nested soft-collinear subtraction scheme~\cite{Caola:2017dug} to
describe production of two equal-mass back-to-back partons.
Integration over the phase space of unresolved radiation, subject to
constraints dictated by the subtraction scheme, was performed using
reverse unitarity~\cite{Anastasiou:2002yz} that allowed us to map
phase-space integrals onto conventional loop integrals with cut
propagators, and apply standard IBP techniques for the reduction of
the integrands to master integrals.
These master integrals were computed by solving a corresponding system
of differential equations in an $\ep$-homogeneous form.

The resulting subtraction terms provide an essential ingredient for
NNLO calculations featuring massive partons.
We note that it is possible to obtain these integrated subtraction
terms numerically, as it was done, for example, in
Refs.~\cite{Czakon:2011ve,Behring:2019oci}.
Nevertheless, it is usually beneficial to have analytic results
available.
The results presented in this article provide all integrated
double-soft subtraction terms required for a description of
colour-singlet decays into massive fermions.
For the case of heavy-quark pair production, it is also necessary to
consider integrated subtraction terms with one massless and one
massive parton which are not necessarily in a back-to-back kinematics.
We leave this problem for future investigations.

\section*{Acknowledgments}
We wish to thank Arnd Behring for providing numerous numerical checks
and Florian Herren for fruitful discussions.
We would like to thank Kirill Melnikov and Fabrizio Caola for valuable
feedback on the manuscript.
This research is partially supported by by the Deutsche Forschungsgemeinschaft
(DFG, German Research Foundation) under grant 396021762 - TRR 257.

\end{document}